# Observation of a structurally driven, reversible topological phase transition in a distorted square net material

Xian P. Yang[1*], Chia-Hsiu Hsu[2,3], Gokul Acharya[4], Junyi Zhang[5], Md Shafayat Hossain[1], Tyler A. Cochran[1], Bimal Neupane[6], Zi-Jia Cheng[1], Santosh Karki Chhetri[4], Byunghoon Kim[1], Shiyuan Gao[5], Yu-Xiao Jiang[1], Maksim Litskevich[1], Jian Wang[7], Yuanxi Wang[6], Jin Hu[4,8], M. Zahid Hasan[1,9,10†]

[1]Laboratory for Topological Quantum Matter and Advanced Spectroscopy (B7), Department of Physics, Princeton University, Princeton, NJ, USA.
[2]Division of Physics and Applied Physics, School of Physical and Mathematical Sciences, Nanyang Technological University, Singapore.
[3]Quantum Materials Science Unit, Okinawa Institute of Science and Technology (OIST), Okinawa 904-0495, Japan.
[4]Department of Physics, University of Arkansas, Fayetteville, AR, USA
[5]Institute for Quantum Matter and Department of Physics and Astronomy, Johns Hopkins University, Baltimore, MD, USA.
[6]Department of Physics, University of North Texas, Denton, TX, USA.
[7]Department of Chemistry and Biochemistry, Wichita State University, Wichita, KS, USA.
[8]Materials Science and Engineering Program, Institute for Nanoscience and Engineering, University of Arkansas, Fayetteville, AR, USA.
[9]Princeton Institute for Science and Technology of Materials, Princeton University, Princeton, NJ, USA.
[10]Lawrence Berkeley National Laboratory, Berkeley, CA, USA.

[*]xiany@princeton.edu
[†]mzhasan@princeton.edu

**Topological materials hold immense promise for exhibiting exotic quantum phenomena, yet achieving controllable topological phase transitions remains challenging. Here, we demonstrate a structurally driven, reversible topological phase transition in the distorted square net material GdPS, induced via *in situ* potassium dosing. Using angle-resolved photoemission spectroscopy and first principles calculations, we demonstrate a cascade of topological phases in the sub-surface P layer: from a large, topologically trivial band gap to a gapless Dirac cone state with a 2 eV dispersion, and finally to a two-dimensional topological insulator as inferred from theory. This evolution is driven by subtle structural distortions in the first P layer caused by potassium adsorption, which in turn contribute to the band gap closure and topological phase transition. Furthermore, the ability to manipulate the topology of a sub-surface layer in GdPS offers a unique route for exploring and controlling topological states in bulk materials.**

Phase transitions mark an abrupt change of a material's electronic properties, driven by external stimuli like temperature, pressure, or chemical composition. A classic example is Anderson localization with a metal to insulator transition when the randomness of a material goes above the critical level [1]. The emerging field of topological materials has expanded this concept to include topological phase transitions [2][3], in which a material transforms between topological and trivial electronic phases, often passing through a metallic state at the critical point [4][5][6][7][8][9][10][11][12][13][14][15][16][17][18][19][20][21][22]. While previous studies have demonstrated topological phase transitions via bulk chemical substitution [4][5][6] or applied strain [7][8][9][10], these methods often lack reversibility, limiting their potential for practical applications. To achieve dynamic control over topological phases, a reversible transition is highly desirable. *In situ* potassium (K) dosing has emerged as a promising approach to achieve such control. For example, in materials like black phosphorus [17], K deposition generates an electric field that leads to band renormalization via the giant Stark effect. Ultrafast photoexcitation has similarly been shown to induce topological phase transitions through crystal lattice modulation [23]. However, in most of these cases, it remains challenging to disentangle whether the transition is driven primarily by electronic doping or by structural modifications.

In this work, we report a reversible, structurally driven topological phase transition in the quantum material GdPS, induced by *in situ* K deposition. GdPS, with its unique crystal structure comprising two-dimensional (2D) P layers distorted from perfect square nets to cis/trans P chains [24][25][26][27][28][29], offers a unique platform for exploring topological phenomena. Our previous theoretical work [30] demonstrated that this structural distortion destroys the linear band crossing in perfect square net materials, resulting in a large band gap in GdPS. Using a combination of angle-resolved photoemission spectroscopy (ARPES) and first principles calculation, we now show that K dosing reversibly drives a sequence of phases in the first P layer: from a large-gap trivial state to a gapless Dirac cone at the quantum critical point, and finally to a two-dimensional topological insulator after band inversion. Importantly, this transition is accompanied by and caused by a subtle but quantifiable lattice distortion in the first P layer, rather than by electronic doping alone. Unlike conventional Dirac semimetals protected by crystal symmetries [31][32][33], the Dirac fermion in GdPS offers greater flexibility for manipulation.

GdPS is a layered material with S-Gd bilayers sandwiched between the P layers along the *c* axis [Fig. 1(a)]. Unlike the perfect square lattice found in some square net materials [Fig. 1(b)], which can host linear Dirac nodal lines [Fig. 1(d)], the P layers in GdPS exhibit a lattice distortion, forming quasi-one-dimensional (1D) armchair chains along the *a* axis [Fig. 1(c)]. The hypothetical square lattice depicted in Fig. 1(b) possesses C2v symmetry, protecting the Dirac point [Fig. 1(d)]. This is analogous to the C2v symmetry protected Dirac cones found in ZrSiS [27], where the Si atoms form a similar square net. However, this symmetry is broken in GdPS due to the distortion of the P layers into armchair chains. This distortion lowers the crystal symmetry, specifically eliminating one of the mirror planes that protect the Dirac point. As our recent DFT calculations demonstrated [30], the symmetry-protected linear band crossing in the ideal square net lattice is gapped out in GdPS [Fig. 1(e)], leading to a substantial band gap [26]. This is consistent with the increase in the P-P bond angle from 90° in Fig. 1(b) to 100.6° in Fig. 1(c), as determined by single crystal X-ray diffraction, which signifies a tetragonal to orthorhombic structural phase transition [34].

Based on the periodicity of the band structure in the out-of-plane $k_z$ direction [34], 81 eV light probes the Γ-X-Y ($k_z$ = 0) plane in our ARPES measurements. The corresponding data are presented in Figs. 2(a)-2(c). The Fermi surface simply consists of a small pocket at the Y point, with additional features at deeper binding energies in Figs. 2(b)-2(c). Our DFT calculations in Figs. 2(f)-2(h) are qualitatively consistent with ARPES data, capturing most of the band dispersions. The band dispersion map along the Γ-Y-Γ direction in Fig. 2(d) reveals a large band gap at the Y point, as expected from the P lattice distortion. Energy distribution curve (EDC) analysis in Fig. 2(e) yields a gap size of approximately 0.74 eV, confirming the substantial band gap induced by the P lattice distortion.

Having established the key features of the electronic structure in pristine GdPS, we now examine the effect of K deposition. While a simple rigid band shift might be expected due to electron transfer from the K atoms, we observe a dramatic and unusual band gap closure, indicating a more complex interaction in GdPS. Figs. 3(a)-3(g) show the evolution of the band dispersion along Γ-Y-Γ with six K dosing cycles. Each cycle lasts 60 seconds with 6.1 A applied current. Initially, the Y point exhibits a large band gap of ~ 0.7 eV [Fig. 3(a)]. Upon K deposition, the electron and hole pockets gradually approach each other, leading to a diminishing band gap, while the flat band in the middle remains relatively unchanged. With further K dosing, the valence and conduction bands eventually touch and then give rise to a band inversion, as shown in Fig. 3(g).

To monitor K coverage, we track the K 3$p$ core level in each dosing cycle [Fig. 3(h)]. The K peak increases until the emergence of a second K 3$p$ peak, which marks the development of a second K layer on the sample surface, consistent with previous studies [35]. Therefore, the fourth and fifth dosing cycles [Figs. 3(e)-3(f)] correspond roughly to a full monolayer of K coverage [35]. Importantly, this is the dosing level where the electron and hole bands touch, signifying the full closure of the band gap. We define this as the critical dosing level.

To further investigate the band crossing at the critical dosing level, we performed photon energy dependent measurements along the Γ-Y direction. As shown in Fig. 3(i), two linear dispersions, corresponding to the two branches of the Dirac cone at Y point, are observed. The lack of dispersion with photon energies confirms the surface nature of this Dirac cone. Due to variations in the cleaved surfaces of the samples, the bulk flat band can be suppressed, enabling a better visualization of the surface Dirac crossing. We repeated the whole K dosing measurements across many samples and selected one with suppressed intensity of the flat band at the critical dosing level [Fig. 4(a)]. The peak positions exacted from the momentum distribution curves (MDCs) in Fig. 4(b) exhibit a clear linear behavior, confirming the presence of a Dirac crossing induced by K dosing on a remarkably large eV energy scale.

To further characterize the band structure beyond the critical K dosing, we again focus on measurements with a suppressed bulk contribution. To disentangle the inverted electron and hole bands, we utilize incident light with linear vertical (LV) and linear horizontal (LH) polarizations, as shown in Figs. 4(c)-4(d). This polarization dependence allows selective excitation of either the hole band (LV) or the electron band (LH) (details on mirror symmetry-driven linear dichroism are in Supplementary Materials [34]). By extracting the MDC peaks and fitting them linearly, we clearly visualize the band inversion in Fig. 4(e). These results confirm that K dosing induces a complete electronic phase transition at the Y point, progressing from a large band gap to a full band closure and finally to a band inversion.

Given the surface origin of the bands at the Y point after K dosing [Fig. 3(i)], we constructed a slab model on the SGd termination in Fig. S2 to understand the evolution of the electronic band structure. This model clearly indicates that ARPES dispersions before K dosing are dominated by the first P layer from layer dependent slab calculations, rather than the top SGd layer above it [34]. Therefore, we focus on the first P layer to model K dosing. Before K deposition in Fig. 4(g), the slab calculation reveals three bands at the Y point: the electron band, the hole band, and the flat band in the middle. These bands agree well with ARPES data, though a rigid downward shift is required to match the experimental Fermi level. This Fermi level offset is a common adjustment in calculations to match ARPES and likely stems from donor-like defects on the sample surface. Nevertheless, the initial phase, prior to K dosing, remains topologically trivial, irrespective of the Fermi level adjustment. Following K dosing, the slab calculations, based on Fig. 4(f), reveal a band inversion near the Y point for the first P layer, while bulk states show minimal change in Fig. 4(h). Consequently, beyond critical K dosing, the first P layer exhibits a band inversion near the Y point, consistent with ARPES data in Fig. 4(e). While there should be a nontrivial band gap arising from the crossing of the band inversion, the gap is so small that ARPES lacks the energy resolution to resolve it and the associated edge state. However, as will be explained later, our DFT clearly reveals both features, confirming its topological nature. At the same time, the bulk bands retain their gap at the Y point, preserving a topologically trivial phase throughout K dosing.

In GdPS, the topological phase transition primarily occurs in the first P layer, with minimal influence from bulk bands. This is supported by slab calculations in Figs. 4(g)-4(h), which show that only the bands from the topmost P layer are significantly affected by K adsorption. To highlight the dominant role of this first P layer, we construct a surface-only model in Fig. S3(b) [34] by extracting the top few atomic layers from the fully relaxed slab structure in Fig. 4(f). This simplified monolayer model [Fig. 4(i)] accurately reproduces the band structure obtained from the full slab calculations in Fig. 4(h). It unequivocally demonstrates that the first P layer undergoes band inversion after K deposition, with $Z_2 = 0$ (1) before (after) K dosing [34]. Furthermore, both the monolayer model in Fig. S3 and the slab model in Fig. 4(f) yield identical $M_{100}$ mirror symmetry eigenvalues for the valence and conduction bands forming the Dirac cone at the Y point (Fig. S4). This underscores the preeminent role of the first P layer in the observed electronic structure.

The monolayer model is subsequently used to investigate the topology of the first P layer in GdPS beyond the critical K dosing. Fig. 4(j) displays the local density of states of the edge states within the 2D band gap of the first P layer. These edge states clearly connect the valence and conduction bands [summarized in Fig. 4(m)], signifying the formation of a topological insulator phase within this layer. In contrast, the bulk band gap persists throughout the K dosing process [grey bands in Figs. 4(g)-4(h)], maintaining a topologically trivial bulk phase [34]. In summary, K dosing induces an electronic phase transition in the first P layer of GdPS, as illustrated in Figs. 4(k)-4(m): the system evolves from a large ~ eV trivial gap in Fig. 4(k) to a gapless Dirac cone at the quantum critical point with a dispersion exceeding 2 eV in Fig. 4(l), and finally to a re-opened nontrivial gap forming a 2D topological insulator in Fig. 4(m).

While the Stark effect can drive topological transitions in other materials [17], the fundamental driving force for the topological phase transition in GdPS is the structural distortion of the first P layer due to K adsorption. To demonstrate this, we calculate two scenarios in Fig. S5 [34]: 1. K dosing without structural relaxation, and 2. the slab first relaxed with K atoms on the surface and then K atoms removed while retaining only the K-induced structural distortion. In the first case, the first P layer's bands shift downwards without any gap reduction, and bulk states remain unchanged. Therefore, if the structure of the first P layer doesn't change, no topological phase transition would occur. This confirms that electron doping alone does not lead to band inversion. In the second scenario, with the K-induced structural distortion retained but without K atoms, the first P layer still exhibits a band inversion. This indicates that it is the structural change on the first P layer that leads to the topological phase transition, directly proving that the observed topological phase transition is structurally driven, not electronically. Furthermore, the P-P bonding angle in the first P layer decreases from 100.5° to 98.0° with K dosing (Fig. S6) [34]. This indicates that K deposition induces a structural change in the top P layer, effectively reducing lattice distortion towards a perfect square net and thereby driving the topological phase transition. Moreover, by continuously tuning the lattice structure with increasing K doping, a Dirac crossing is formed at the quantum critical point [Fig. 4(a)], consistent with previous studies [36][37][38].

Beyond theoretical calculations, we provide direct experimental evidence of a subtle lattice modification in the first P layer. ARPES is highly sensitive to this layer, allowing us to track its structural changes precisely. We quantify the change by measuring the distance between the centers of the two Y pockets in the first Brillouin zone. Figure S7 presents momentum distribution curves (MDCs) along the Y-Γ-Y direction, extracted before and at the critical K dosing level across multiple binding energies [34]. Remarkably, the separation between the two Y points decreases systematically after K dosing, providing direct momentum-space evidence of an in-plane lattice change in the first P layer.

Crucially, the K dosing process and the accompanying structural changes are reversible upon heating. Heating the sample removes K atoms from the surface and restores the pristine electronic structure. Experimentally, we demonstrate this reversibility by heating the sample at the critical dosing level [Fig. S9(b)] [34], which fully restores the large topologically trivial gap [Fig. S9(c)]. Therefore, the topological phase transition in GdPS can be precisely and reversibly controlled through K deposition and desorption [Figs. 4(k)-4(m)].

It may be tempting to attribute the Dirac crossing at the critical dosing level to a full restoration of the mirror symmetry broken in pristine GdPS due to the distorted square lattice [30]. If this were the case, the corresponding Fermi surface should exhibit fourfold symmetry consistent with a restored tetragonal square net. However, our ARPES data at the critical dosing still shows a twofold rotational symmetry [34], identical to pristine GdPS [Fig. 2(a)]. This indicates that, while the structurally driven gap closure reduces lattice distortion, it does not fully restore a perfect square net lattice. This is consistent with DFT calculations, which show that the P-P bonding angle, although reduced, remain above 90° at the critical dosing. Consequently, the K-dosing induced

Dirac dispersion represents a novel phase distinct from the linear dispersion in ideal square net lattices. The fact that such a subtle P-P lattice angle change can drive a dramatic topological phase transition highlights the sensitivity of electronic and topological properties to the underlying lattice structure. This contrasts with LaSb$_x$Te$_{2-x}$ [22], where glide symmetry is restored after K deposition.

Although the topological phase transition occurs in a 2D P layer within a 3D material, the first P layer in GdPS is beneath the SGd surface layer rather than the topmost cleaved surface. We confirmed this by comparing experimental ARPES data and termination-dependent slab calculations, where only the SGd termination matches reasonably well with experimental results [34]. Given that ARPES signals are dominated by the contribution from the first P layer under the SGd layer, GdPS offers a rare opportunity to visualize and control the topological phases underneath the surface of the sample. It is remarkable that surface K adsorption can induce structural and topological changes in a buried P layer. This unique scenario highlights the potential for manipulating the electronic properties below the sample surface in bulk crystals.

In conclusion, our study reveals a structurally induced, reversible topological phase transition in GdPS. Starting from a large (~1 eV) trivial band gap, K dosing drives the system through a gapless Dirac cone with a large (>2 eV) linear dispersion at the critical dosing, and finally to a 2D topological insulator. Our results show that K dosing induces a structural distortion in the top P layer, and it is this structural change, rather than electron doping alone, that drives the topological phase transition in GdPS. Compared to previous studies [17], this work distinguishes itself by combining several key aspects within a new material system. Firstly, the K-induced topological phase transition in GdPS is fully reversible and structurally driven, allowing for dynamic and precise control of the electronic properties. Moreover, our work demonstrates the remarkable ability to manipulate the topology of a sub-surface layer within a bulk crystal. Our work thereby identifies GdPS as a promising platform for exploring and controlling topological states in bulk materials, with potential applications in novel devices exploiting tunable topological phase transitions.

**Acknowledgement**. We acknowledge illuminating discussions with Guoqing Chang and Titus Neupert. Advanced ARPES and theoretical work at Princeton University were supported by the US DOE under the Basic Energy Sciences program (grant number DOE/BES DE-FG-02-05ER46200; M.Z.H.), National Quantum Information Science Research Centers, Quantum Science Center (at ORNL) supported by US DOE and the Gordon and Betty Moore Foundation (GBMF9461; M.Z.H.). This research used resources of the Advanced Light Source, which is a DOE Office of Science User

Facility under contract no. DE-AC02-05CH11231. We acknowledge Diamond Light Source for time on beamline I05. We also acknowledge the Paul Scherrer Institut, Villigen, Switzerland for provision of synchrotron radiation beamtime at the Surface/Interface Spectroscopy (SIS) beamline of the SLS. This research used the ESM beamline of the National Synchrotron Light Source II, a U.S. Department of Energy (DOE) Office of Science User Facility operated for the DOE Office of Science by Brookhaven National Laboratory under Contract No. DE-SC0012704. The authors want to thank S.-K. Mo at Beamline 10.0.1 of the ALS, C. Cacho and T. Kim at beamline I05 of Diamond Light Source, and N.C. Plumb and Y. Hu at SIS beamline of the SLS for support in getting the ARPES data. The authors also want to thank E. Vescovo, A. Rajapitamahuni and T. Yilmaz at Beamline 21-ID-1 (ESM-ARPES) of the National Synchrotron Light Source II for support in getting the preliminary ARPES data. J.H acknowledges the support by the U.S. Department of Energy, Office of Science, Basic Energy Sciences program under Grant No. DE-SC0022006 for crystal growth. B.N. and Y.W. acknowledge startup funds from the University of North Texas, and computational resources from the Texas Advanced Computing Center. Part of the modeling was supported by computational resources from a user project at the Center for Nanophase Materials Sciences (CNMS), a US Department of Energy, Office of Science User Facility at Oak Ridge National Laboratory, and also partially by user project R0076 at the Pennsylvania State University Two-Dimensional Crystal Consortium – Materials Innovation Platform (2DCC-MIP) under NSF cooperative agreement DMR-2039351. JYZ acknowledges support from the NSF CAREER Grant DMR-1848349, Johns Hopkins University Theoretical Interdisciplinary Physics and Astronomy Center and the Institute for Quantum Matter, an Energy Frontier Research Center funded by the U.S. Department of Energy, Office of Science, Office of Basic Energy Sciences, under Award DESC0019331. Work at Wichita (single crystal XRD and structure refinement) was supported by NSF under award OSI-2328822. M.Z.H. acknowledges support from Lawrence Berkeley National Laboratory and the Miller Institute of Basic Research in Science at the University of California, Berkeley in the form of a Visiting Miller Professorship. M.Z.H. also acknowledges support from the U.S. Department of Energy, Office of Science, National Quantum Information Science Research Centers (ORNL), Quantum Science Center and Princeton University.

X. P. Y., C.-H. H., and G. A. contributed equally to this work.

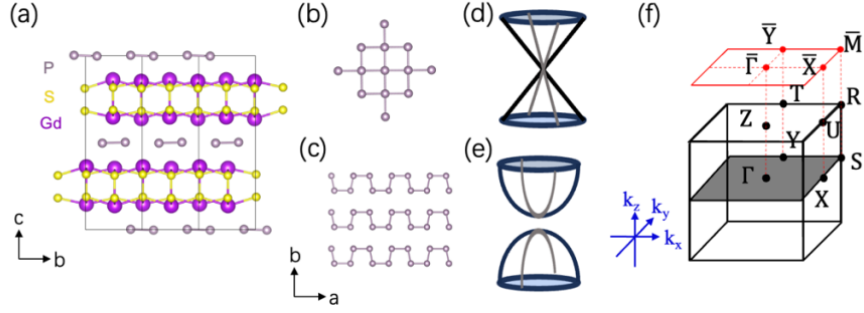

FIG. 1. P lattice distortion in GdPS. (a) Crystal structure of GdPS. (b) Square net lattice. (c) The P layer in GdPS is distorted from the perfect square net in (b). (d) The linear Dirac like crossing from square net lattice in (b). (e) The P lattice distortion gaps out the band crossing. (f) Bulk and surface Brillouin zones of GdPS.

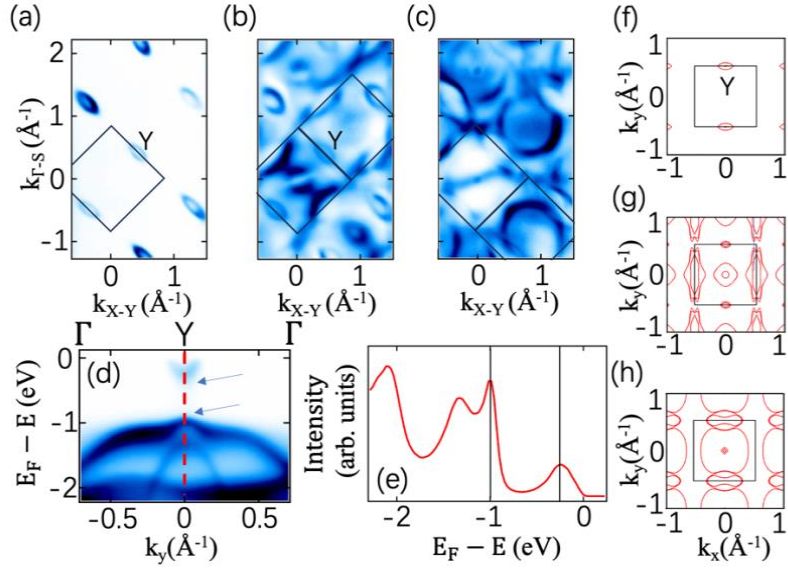

FIG. 2. ARPES band dispersion of GdPS. (a) ARPES Fermi surface spectrum on the Γ-X-Y high symmetry plane. Black box indicates the Brillouin zone. There is a small electron pocket at the Y point. (b-c) ARPES constant energy contours at binding energies of 1.5 and 2.1 eV. (d) ARPES dispersion map along the Γ-Y-Γ direction demonstrating a large band gap at the Y point. (e) Energy distribution curve (EDC) marked as the red dotted line in (d). Two peaks indicated by the black vertical lines represent the energy position of the electron pocket and the flat hole band also marked by the blue arrows in (d). The band gap between these two bands is extracted to be 0.74 eV. (f-h) Calculated constant energy contours corresponding to ARPES spectra in (a-c). The black box represents the Brillouin zone and the location of the Y point is marked.

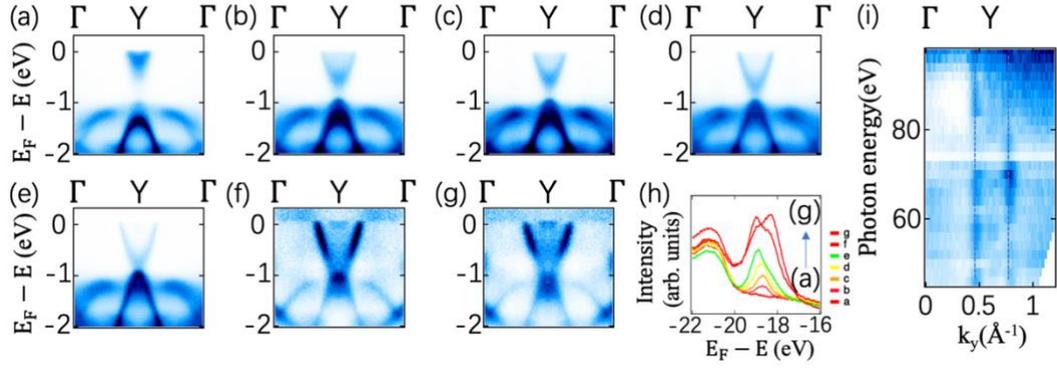

FIG. 3. K dosing induced electronic phase transition at the Y point. (a) ARPES dispersion map along the Γ-Y-Γ direction before K dosing. (b-g) ARPES dispersion maps along the Γ-Y-Γ direction taken after repeated K dosing cycles. Each dosing cycle takes one minute. (h) K 3*p* core level spectra for all dosing cycles, corresponding to (a-g) from bottom to top. Each letter represents a distinct dosing cycle. (i) Photon energy dependence along the Γ-Y direction. Vertical blue dotted lines confirm that the two branches of the linear crossing at Y point have a surface nature.

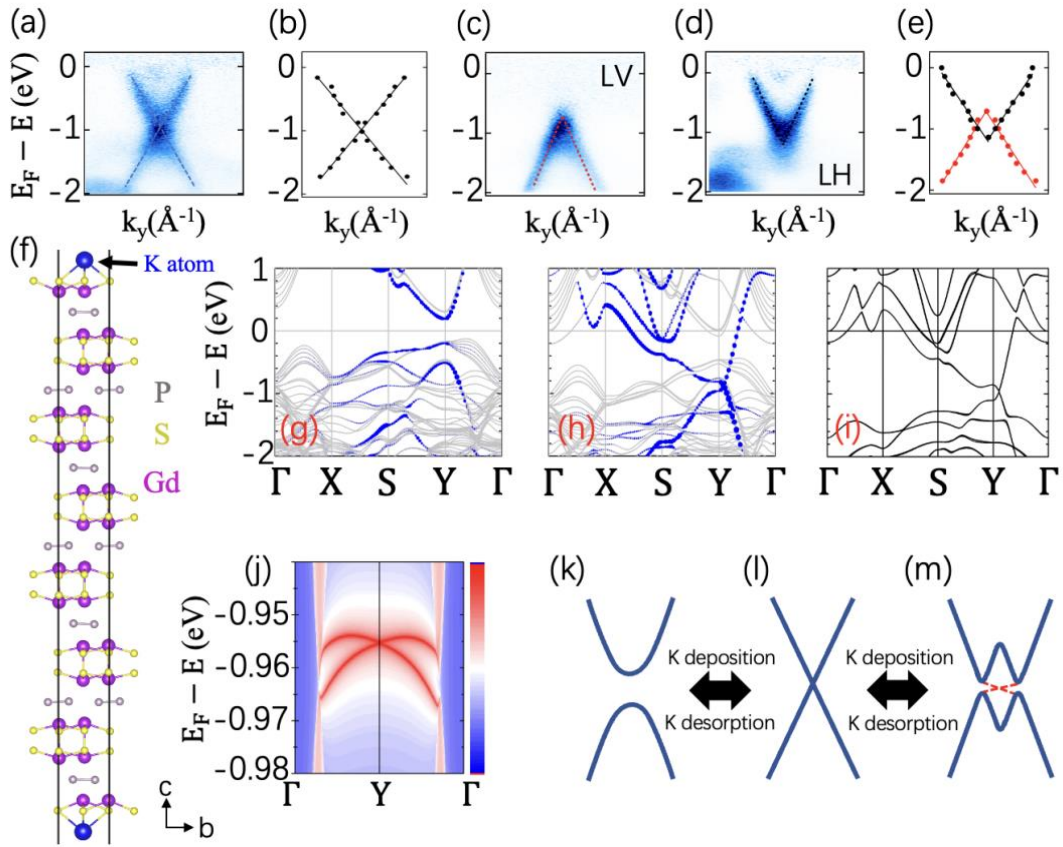

FIG. 4. K dosing driven topological phase transition on the surface of GdPS. (a) ARPES dispersion map at the critical dosing level highlighting the linear crossing at the Y point. Blue dotted lines are a guide for the eye. The flat hole band is suppressed in this measurement. (b) Black dots are the extracted peak positions corresponding to the momentum distribution curves (MDCs) of the linear crossing in (a). Black lines are the linear fits. (c-d) ARPES dispersion map beyond the critical dosing

level at the Y point taken with LV [(c)] and LH [(d)] incident lights. Black and red dotted lines are the linear fits to the extracted MDC peaks in (e). (e) Black and red dots are the extracted MDC peak positions based on the linear bands in (c) and (d), respectively. Black and red lines are the corresponding linear fits. (f) A slab model of SGd termination consisting of six S-Gd bilayers. The position of the deposited K atom is marked by the black arrow. (g) DFT calculation of the band structure before K dosing based on the slab model in Fig. S2(a). The blue circles denote the contribution of the first P layer, while the grey bands are from the bulk. (h) DFT band dispersion beyond the critical dosing level based on the slab model in (f). The blue circles represent the contribution from the first P layer, and the grey bands are bulk states. (i) DFT calculation based on the monolayer model in Fig. S3(b). (j) Zoom-in plot of (i) demonstrating the edge states (red dispersions) inside the 2D band gap in the first P layer. (k-m) Simplified schematics indicating the reversible topological phase transition controlled by K deposition and desorption. The edge states in (m) are marked by the red dotted line.

**Supplementary Materials**

**ARPES.** ARPES measurements were conducted at the ULTRA endstation of the Surface/Interface Spectroscopy (SIS) beamline of the Swiss Light Source, Beamline 10.0.1 of the Advanced Light Source (ALS) in Berkeley, California, USA, as well as the I05 beamline of Diamond Light Source in UK. GdPS samples were cleaved *in situ* with a pressure better than $5 \times 10^{-11}$ torr. The energy and angle resolutions were better than 30 meV and 0.2°, respectively. Incident photon energy was set to 81 eV for most of the measurements and temperature was kept at 20 K at ULTRA and I05 beamline but at 90 K at ALS. The Fermi level was determined by measuring Au that is in electrical contact with the samples. Potassium atoms were evaporated *in situ* on GdPS samples using a commercial SAES potassium dispenser. Each dosing cycle took 60 seconds with 6.1 A applied current. To remove K on the sample surface, we heated the samples at around 100 °C for 30 min. The sample was then transferred back to the main chamber for ARPES measurement.

We note that the change in the momentum separation between the two Y points directly reflects a modification of the in-plane lattice constant, since the reciprocal space distance is inversely proportional to the real space periodicity. Fig. S7 shows that this separation shifts systematically upon K dosing, providing experimental evidence for a surface structural change. To ensure that this shift is intrinsic rather than caused by angular drift or instrumental artifacts, Fig. S8 compares the surface bands with bulk bands at a deeper binding energy. While the surface bands exhibit a measurable k-shift, the bulk bands remain at fixed momentum positions within experimental resolution, serving as an internal angular reference. Modern ARPES instruments maintain a stable and calibrated electron emission geometry, and the unchanged bulk band separation confirms that the observed shift originates from the sample rather than extrinsic factors.

**Sample preparation.** Single crystals of GdPS were grown by a direct chemical vapor transport with a stoichiometric ratio of Gd, P, and S as source materials and $TeCl_4$ as the transport agent. The growths were performed in a dual heating zone furnace with a temperature gradient from 1075 to 975 °C for three weeks. The structure of single crystal GdPS samples was determined by single crystal XRD.

**First principles calculations.** Based on the density functional theory (DFT) framework [39], the first-principles calculations were performed using the Vienna ab-initio simulation package (VASP) [40][41] with the projector-augmented-wave (PAW) potentials [42] under generalized gradient approximation of the Perdew-Burke-Ernzerhof (PBE) [43]. The kinetic energy cutoff was set at 300 eV. Structural relaxation was performed until the residual forces were no greater than $10^{-3}(10^{-2})$ eV/Å, and the energy convergence criteria for self-consistency were set at $10^{-6}(10^{-4})$ eV for the bulk (slab) system. Γ-centered Monkhorst-Pack grids of $12 \times 12 \times 4$ and $6 \times 6 \times 1$ in the first Brillouin zone (BZ) were used for the bulk and slab systems, respectively. To avoid the interaction between the adjacent layers due to periodic boundary conditions, a vacuum of ~20 Å was added for the slab system. The $Z_2$ invariants with Wannier center charge which is used for the system without inversion symmetry [44] and edge state using surface Green's function [45] within the tight-binding framework were calculated to identify the topological properties of the monolayer system using the WannierTools program [46] based on the maximally localized Wannier functions (MLWF) obtained from the Wannier90 package [47]. The Irvsp package was used to identify the symmetry characteristics of bands by calculating the eigenvalues (trace) of the mirror symmetry operator [48].

We note that the lattice response to K dosing is strongly localized at the surface. As shown in Fig. S6, only the first P layer exhibits a significant change in in-plane lattice constant, while the second and deeper layers remain essentially unchanged. The topmost SGd layer undergoes a moderate structural adjustment but the distortion rapidly decays with depth. Therefore, although a small lattice mismatch may occur between the first P layer and the SGd layer, the structural modification is confined to the near-surface region, and the development of long-range superperiodicity in the bulk is unlikely.

In GdPS, the structural change occurs concurrently with electron doping from K atoms, so the two effects cannot be experimentally separated. However, isolating them is not the central goal of this study. The key finding is that surface doping induces an unexpected structural distortion, which in turn drives the topological phase transition. This mechanism is fundamentally different from prior work in black phosphorus, where excess electron doping alone drives a purely electronic transition. In GdPS, it is the combination of electron doping and the resulting structural modification that triggers the topological transition. The experimental evidence in Fig. S7, together with our DFT calculations, sufficiently demonstrates this mechanism. While *in situ* strain control could be explored in future work, it is not required to support the conclusions presented here.

Finally, the 2D topological insulator (2D TI) phase discussed in this work is inferred from DFT calculations. The small size of the inverted gap (~meV) prevents direct experimental resolution by ARPES, and the associated edge states cannot be probed via STM due to the surface K atoms covering the P layer. However, our experimental data clearly show the band inversion after K dosing (Figs. 4(c)–(e)), which provides indirect, complementary support for the theoretical prediction. The central novelty of this study lies in demonstrating that K dosing induces a surface structural change that drives the topological phase transition, rather than in the direct experimental observation of the 2D TI edge states.

**Termination dependent slab calculations.** There are three terminations in GdPS: P, GdS and SGd terminations. In Fig. S11, we show the band structure dispersions for P and GdS terminations before K dosing. However, both the top and second layers for both terminations don't show the band dispersions in ARPES [Figs. 2 and 3]. Fig. S2 demonstrates the band structure for SGd termination before K dosing, which matches well with ARPES results in Fig. 2. In other words, since both GdS and P terminations don't show the dispersions observed in ARPES, the experimental cleavage plane should be the SGd termination. Different from P-termination where the top surface is the first P layer, the top surface of SGd termination is the SGd layer, and the first P layer is the second layer [for example, Fig. 4(f)]. Given that ARPES signals are dominated by the contribution from the first P layer under the SGd layer [as shown in Figs. 4(g)-4(h)], GdPS offers a rare opportunity to visualize and control the topological phases underneath the surface of the sample.

**$M_{100}$ mirror symmetry driven linear dichroism.** Beyond the critical dosing level in GdPS, a strong linear dichroism is observed in ARPES [Figs. 4(c)-4(d)]. Linearly vertical (LV) polarized light selectively excites the hole band, while linearly horizontal (LH) polarized light excites the electron band. This allows us to isolate and visualize each band individually. This linear dichroism arises from the $M_{100}$ mirror symmetry, which protects the band inversion at the Y point.

The Dirac bands at the Y point are protected by the $M_{100}$ mirror symmetry (y-z plane), perpendicular to the 1D armchair P chains. Eigenvalue (trace) of $M_{100}$, calculated using the Irvsp package near the Dirac point [Fig. S4(b)], reveals opposite eigenvalues (+1 and -1) for the bands around the Y point. Moreover, band structure calculations in Fig. S4(d) based on a monolayer model [Fig. S3(b)] after K dosing confirm that the bands around the Y point have the same symmetry character as the slab model in Fig. 4(f) [based on the comparison between Figs. 4(b) and S4(d)]. Therefore, the monolayer model is used to investigate the nontrivial topology of the first P layer beyond the critical K dosing in Fig. 4(j). In summary, the $M_{100}$ eigenvalues for the Dirac cone bands are opposite (+1 and -1).

ARPES intensity is governed by the matrix element effect *M* (or selection rule), which reflects the orbital symmetry of the bands. Nonzero ARPES intensity requires a nonzero transition matrix *M* ~< *I* |*A*\**p*| *f* > relative to specific mirror planes, where *A* is the electric field vector and *p* is the momentum operator. Under normal emission, the final state |*f*> has even symmetry with respect to the GdPS surface mirror plane. Thus, orbitals (or the initial state < *i* |) with even (odd) symmetry are detected by incident light with even (odd) symmetry (or electric field vector *A*) with respect to the same mirror plane.

The experimental geometry in Fig. S4(a) highlights the $M_{100}$ mirror symmetry protecting the band inversion at the Y point. In our experiment, LV polarized light is along the *x* axis, and LH light is within the $M_{100}$ mirror plane in Fig. S4(a). Therefore, LH and LV have even and odd symmetries relative to $M_{100}$, respectively. Since the conduction and valence bands at the Y point have $M_{100}$ eigenvalues of 1 and -1 [Fig. S4(b)], their initial states are even and odd, respectively. Consequently, LH light with even symmetry detects the conduction band, and LV light with odd symmetry detects the valence band, explaining the observed linear dichroism in Figs. 4(c)-4(d).

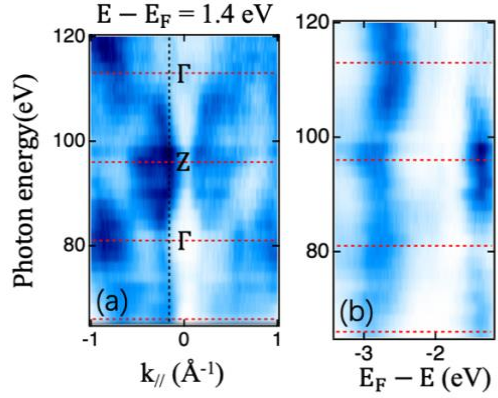

FIG. S1. Photon energy dependence along $\bar{\Gamma}$-$\bar{M}$ direction. (a) Photon energy dependence along the $\bar{\Gamma}$-$\bar{M}$ direction. The constant energy contour is at binding energy of 1.4 eV. Red dotted lines mark the position of high symmetry points, and the black dotted line indicates the location of the ARPES dispersion map in (b). (b) ARPES dispersion map corresponding to the black dotted line in (a). From the periodicity of the band structure in the $k_z$ direction, 81 eV incident light probes the $\Gamma$-X-Y plane.

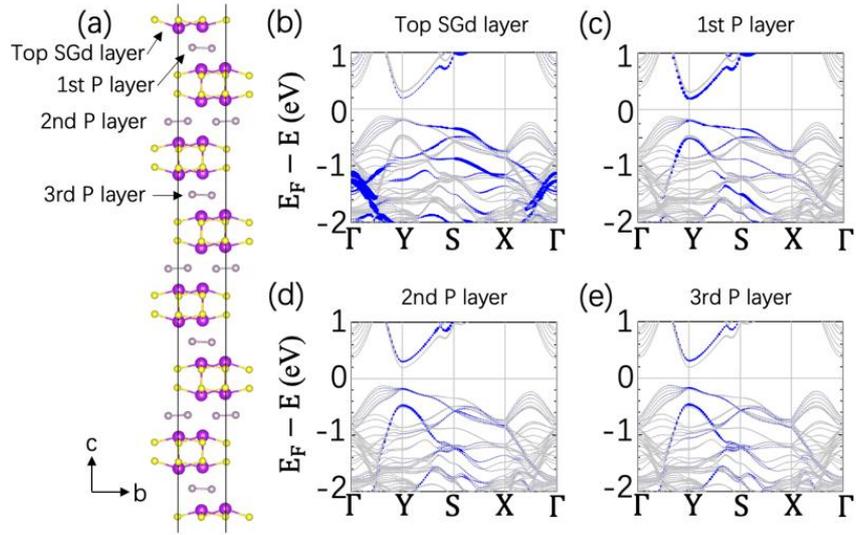

FIG. S2. Layer dependent slab calculation on the SGd termination before K dosing. (a) A slab model on the SGd termination consisting of six S-Gd bilayers before K dosing. (b-e) DFT calculation of the band structures (grey lines) before K dosing based on the slab model in (a). The blue circles denote the contribution from the top SGd layer in (b), first P layer in (c), second P layer in (d) and third P layer in (e). A comparison between (b) and (c) indicates the ARPES signals before K dosing are dominated by the first P layer.

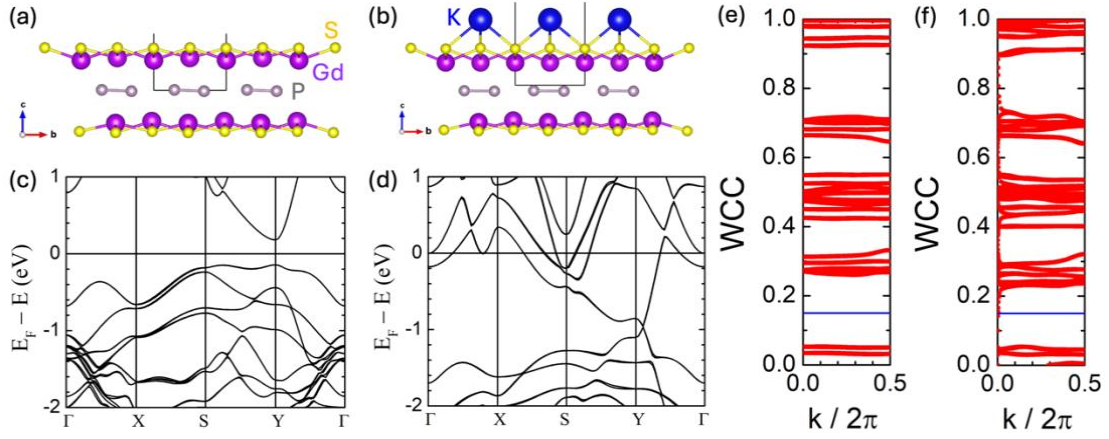

FIG. S3. Monolayer calculation before and after K dosing. (a-b) A monolayer model cleaved from the top GdPS layer of the slab without and with K dosing. (c-d) DFT calculations based on the monolayer models in (a-b). (e-f) The evolutions of Wannier charge centers on the $k_z = 0$ plane before (e) and after (f) K dosing for the monolayer models in (a-b). The evolution curves cross any arbitrary reference line (like the blue dotted lines) an even (old) number of times, yielding $Z_2 = 0$ (1). Therefore, K dosing induces a topological phase transition from a trivial phase to a nontrivial phase.

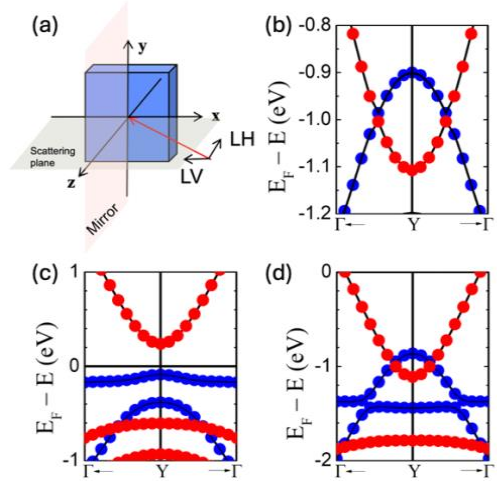

FIG. S4. (a) Experimental geometry of polarization dependent ARPES. GdPS sample denoted by blue cuboid lies within the *x-y* plane. Red arrow indicates the incident light and black arrows mark the direction of LV and LH polarization. Scattering plane under normal emission and mirror plane are also shown. (b) The band structure of the first P layer beyond the critical dosing level with the eigenvalue of the $M_{100}$ symmetry operator around the Y point for the slab model in Fig. 4(f). The red (blue) circles denote eigenvalue of $+1$ $(-1)$ on the conduction (valence) band. (c-d) The band structure before and after K dosing with the eigenvalue of the $M_{100}$ symmetry operator around the Y point based on the monolayer models in Figs. S3(a)-3(b). The red (blue) circles denote eigenvalue of $+1$ $(-1)$ on the conduction (valence) band. A comparison between (b) and (d) confirms the bands around the Y point have the same symmetry character for the monolayer model as well as the slab model. Therefore, the monolayer model can be applied to visualize the nontrivial topology in the first P layer beyond the critical K dosing.

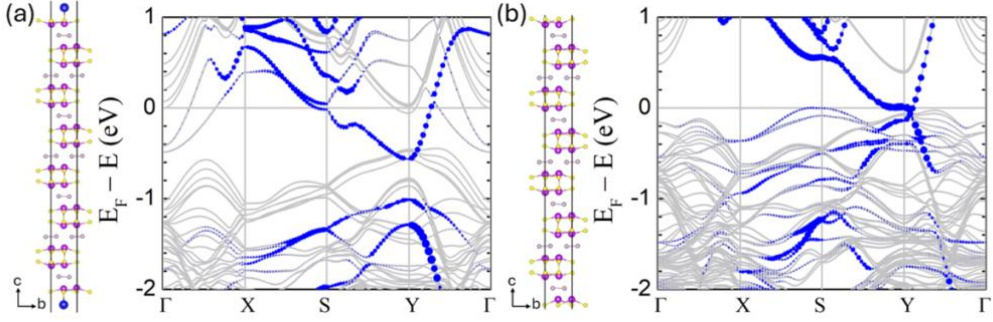

FIG. S5. Band structures of two artificial slab models constructed to distinguish the effects of K deposition. (a) K atom is placed on the surface of the GdPS slab without structural relaxation acting as pure electron doping. (b) Structure is relaxed with the K atom on the surface of the slab, and then the K atom is removed capturing K-induced structural distortion only. The blue circles indicate the contribution from the first P layer. The corresponding P-P bond angles of the first P layer are 100.8° and 97.99° for (a) and (b), respectively. These models allow us to separate the effects of electron doping and surface structural modification on the band structure.

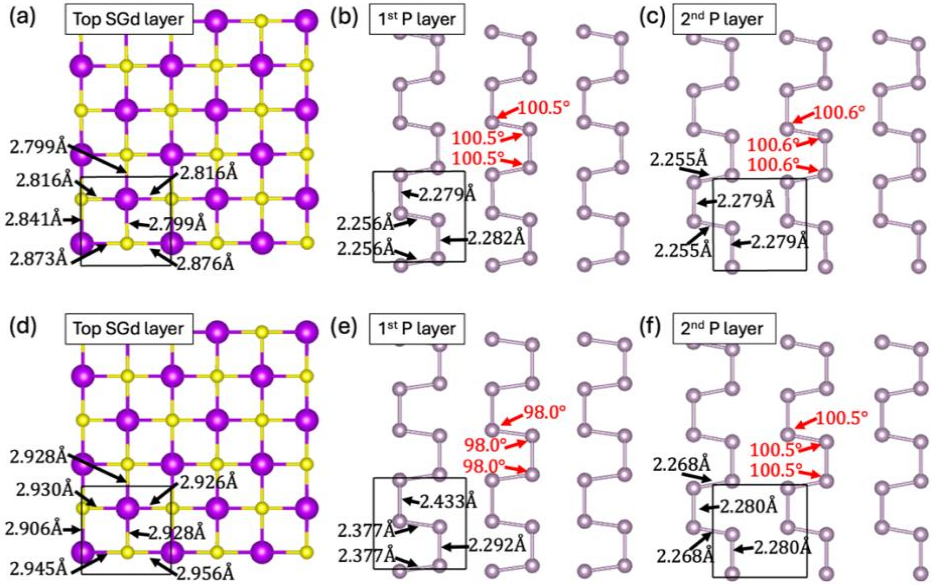

FIG. S6. DFT calculations revealing the structural changes of the first P layer on the SGd termination. (a-c) The structure of top SGd layer (a), 1st P layer (b) and 2nd P layer (c) on the SGd termination before K dosing. (d-f) The structure of top SGd layer (d), 1st P layer (e) and 2nd P layer (f) on the SGd termination after K dosing. A comparison between (b) and (e) clearly demonstrates the change of P-P bonding angle on the first P layer after K dosing, while the second P layer has negligible structural change [(c) and (f)].

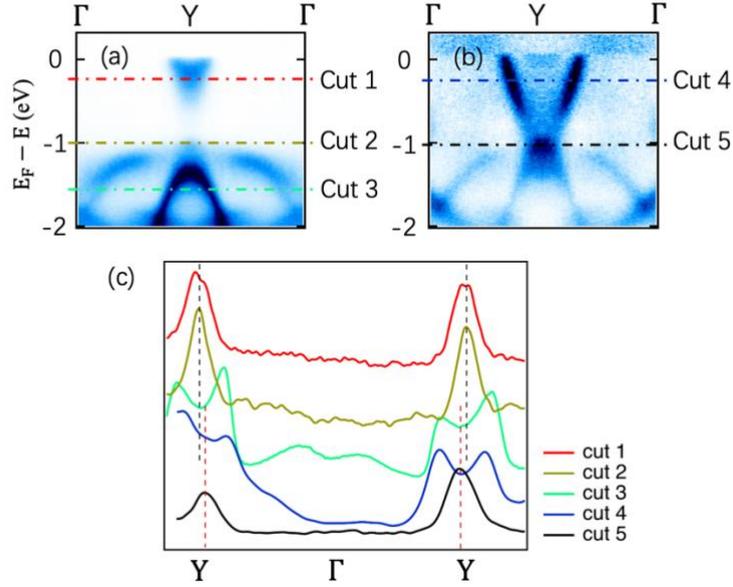

FIG. S7. Evidence of a subtle lattice change on the first P layer after K deposition. (a) ARPES dispersion map along the Γ-Y-Γ direction before K dosing. Same as Fig. 3(a). (b) ARPES dispersion map along the Γ-Y-Γ direction at the critical K dosing level. Same as Fig. 3(f). (c) Five curves (cuts 1-5 from top to bottom) represent the momentum distribution curves (MDC) taken at various binding energies for before (cuts 1-3 in (a)) and at critical K dosing (cuts 4-5 (b)), respectively. Black and red vertical dotted lines represent the momentum position of the Y point, which demonstrate a slight decrease after K deposition confirming the lattice change along the Y-Γ direction.

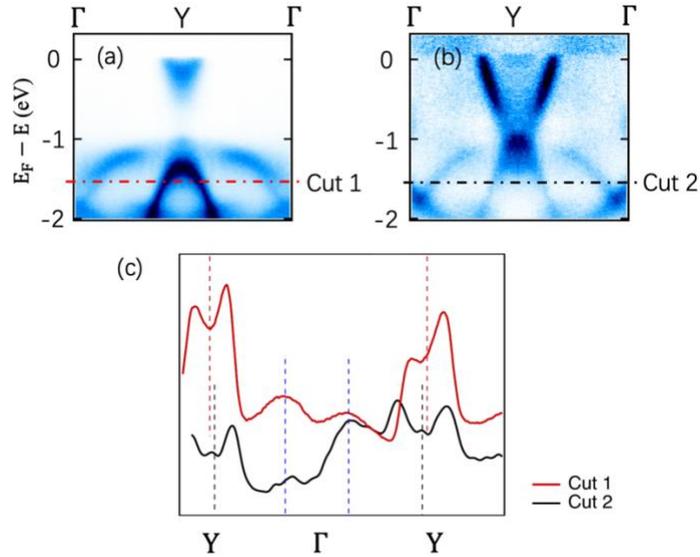

FIG. S8. Angular calibration confirming that only the surface bands shift after K deposition. (a) ARPES dispersion map along the Γ-Y-Γ direction before K dosing. Same as Fig. 3(a). Cut 1 goes through both surface and bulk bands near Y and Γ points, respectively. (b) ARPES dispersion map along the Γ-Y-Γ direction at the critical K dosing level. Same as Fig. 3(f). Cut 2 goes through both surface and bulk bands near Y and Γ points, respectively. (c) MDCs for before (cut 1 in (a)) and at critical K dosing (cut 2 in (b)), respectively. Red and black vertical dotted lines represent the momentum position of the Y point, which demonstrate a slight decrease after K deposition confirming the lattice change on the first P layer along the Y-Γ direction. Blue vertical dotted lines mark the momentum positions of the bulk band near the Γ point, which remain unchanged within experimental resolution after K dosing. This demonstrates that the observed k-shift in Fig. S7 is intrinsic to the surface states and not due to instrumental angular drift or extrinsic variations in measurement conditions.

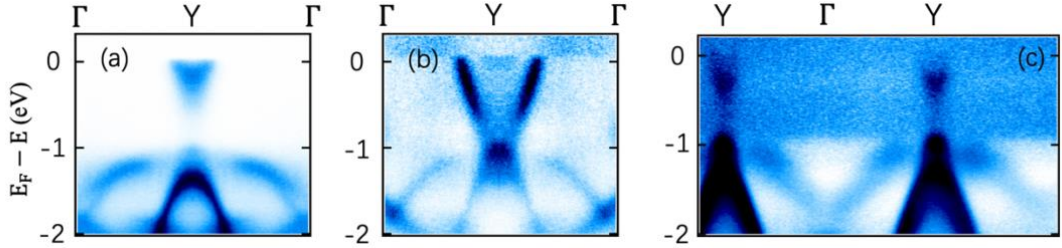

FIG. S9. Reversible topological phase transitions. (a) ARPES dispersion map along the Γ-Y-Γ direction before K dosing. Same as Fig. 3(a). (b) ARPES dispersion map along the Γ-Y-Γ direction at the critical K dosing level. Same as Fig. 3(f). (c) Heating the sample at the critical dosing level in (b) restores the topologically trivial gap in (a).

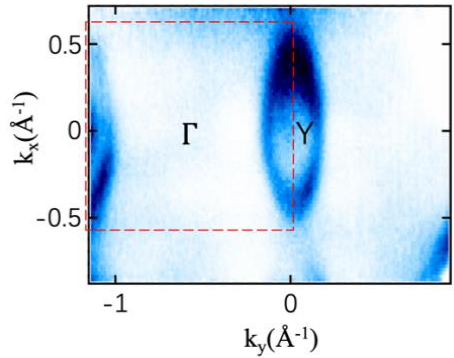

FIG. S10. Fermi surface at the critical dosing. ARPES Fermi surface spectrum on the Γ-X-Y high symmetry plane at the critical dosing. Red box indicates the Brillouin zone and the high symmetry points are marked.

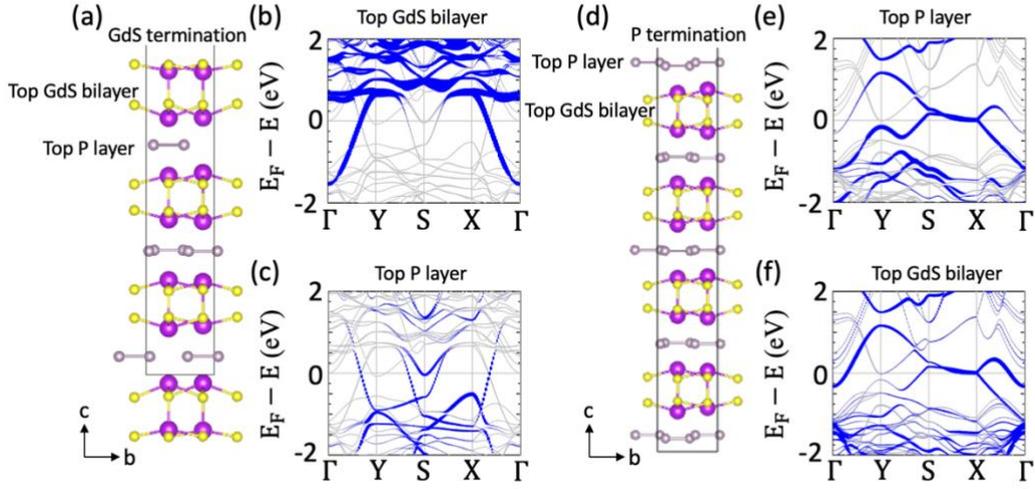

FIG. S11. Termination dependent slab calculations on the SGd and P terminations before K dosing. (a) A slab model on the SGd termination before K dosing. (b-c) DFT calculation of the band structures (grey lines) before K dosing based on the slab model in (a). The blue circles denote the contribution from the top GdS bilayer in (b) and the first P layer in (c). (d) A slab model on the P termination before K dosing. (e-f) DFT calculation of the band structures (grey lines) before K dosing based on the slab model in (d). The blue circles denote the contribution from the top P layer in (e) and the first GdS bilayer in (f). A comparison between Fig. S11 and Fig. 2 confirm ARPES observes the SGd termination in Fig. S2, since only the SGd termination matches ARPES results reasonably well.